\documentclass[conference]{IEEEtran}
\IEEEoverridecommandlockouts
\usepackage{cite}
\usepackage{amsmath,amssymb,amsfonts}
\usepackage{algorithmic}
\usepackage{graphicx}
\usepackage{textcomp}
\usepackage{xcolor}
\usepackage{wrapfig}
\usepackage{svg}

\def\BibTeX{{\rm B\kern-.05em{\sc i\kern-.025em b}\kern-.08em
    T\kern-.1667em\lower.7ex\hbox{E}\kern-.125emX}}

\usepackage[acronym]{glossaries}

\newacronym{bft}{BFT}{Byzantine Fault Tolerant}
\newacronym{cft}{CFT}{Crash Fault Tolerant}
\newacronym{dt}{DT}{Digital Twin}
\newacronym{soc}{SOC}{Security Operations Center}
\newacronym{cps}{CPS}{Cyber-Physical Systems}
\newacronym{tsn}{TSN}{Time-Sensitive Networking}
\newacronym{tsn2}{TSN/L2}{Time-Sensitive Networking at Layer 2}
\newacronym{opc}{OPC UA}{Open Platform Communications Unified Architecture}
\newacronym{vlan}{VLAN/TSN}{Virtual LAN segmentation with Time-Sensitive Networking}
\newacronym{plc}{PLC}{Programmable Logic Controller}
\newacronym{ot}{OT}{Operational Technology}
\newacronym{otw}{OTw}{Operational Twin}
\newacronym{siem}{SIEM}{Security Information and Event Management}
\newacronym{fl}{FL}{Federated Learning}
\newacronym{dp}{DP}{Diferential Privacy}
\newacronym{k8s}{K8s}{Kubernetes}
\newacronym{k3s}{K3s}{Lightweight Kubernetes}
\newacronym{hla}{HLA}{High-Level Architecture}
\newacronym{fmu}{FMU}{Functional Mock-up Unit}

\begin{document}

\title{Trust, but Verify: ByzTwin-Range, a Digital Twin Cyber-Range for Byzantine Faults
}

\author{\IEEEauthorblockN{Tadeu Freitas, João Soares}
\IEEEauthorblockA{Department of Computer Science,\\ Faculty of Science, University of Porto\\Email:\{tadeufreitas, joao.soares\}@fc.up.pt}
\and
\IEEEauthorblockN{Rolando Martins}
\IEEEauthorblockA{SafeHelm, lda,\\ Porto, Portugal\\Email: rmartins@safehelm.com
}
}
\IEEEaftertitletext{\vspace{-20pt}}
\maketitle

\begin{abstract}
Critical infrastructures increasingly rely on interconnected and software-driven \ \gls{cps}, exposing operational processes to both accidental failures and sophisticated adversarial behavior. 

While \gls{bft} protocols offer robustness against arbitrary faults, evaluating their behavior under realistic cyber-physical conditions remains challenging: traditional cyber ranges lack timing fidelity, and testing in production environments is unsafe. 
This paper introduces ByzTwin-Range, a dual-layer architecture that integrates a production-grade \gls{bft} deployment with a \gls{dt} to enable controlled experimentation, stress testing, and Byzantine fault injection using live operational data. 

The \gls{dt} mirrors real system state, executes ``What-if'' analyses through co-simulation and emulation, and identifies synchrony vulnerabilities, i.e., misconfigured timeouts, timing-sensitive false suspicions, and adversarial delay exploits, configuration weaknesses, and adversarial behaviors that may undermine \gls{bft} guarantees. 
Insights from the twin are fed back into the operational deployment through a secure advisory channel, supporting continuous validation and adaptive hardening. The proposed design leverages industry-standard technologies (\glsentrylong{opc}, \glsentrylong{tsn}, \glsentrylong{fmu}/\glsentrylong{hla}, QUIC/mutual TLS) to maximize feasibility and compatibility with existing industrial workflows.

ByzTwin-Range establishes a practical foundation for next-generation, \gls{bft}-aware cyber ranges and paves the way for more resilient \glspl{cps} through continuous testing, differential-privacy-enabled analytics, and future proof-of-concept implementations.
\end{abstract}

\begin{IEEEkeywords}
BFT, Digital Twins, Cyber-Physical Systems, Resilient Control Systems
\end{IEEEkeywords}

\glsresetall

\section{Introduction and Motivation}
In recent years, critical infrastructures have undergone continuous modernization, driven by the integration of advanced IT systems—including sensors, automated detectors, software-driven controllers, and networked supervisory platforms—to improve operational efficiency and quality of service. 
This convergence of digital technology with traditionally isolated operational environments has enabled new levels of automation, visibility, and optimization~\cite{gheorghe2024review}.

However, both historical incidents~\cite{Falco2012StuxnetFacts,Case2016_UkrainePowerGrid,CRS2021_ColonialPipeline} and current trends~\cite{Ashley2022_AggregateAttackSurface, Nankya2023_ICS_Security, ENISA2024_ThreatLandscape} clearly demonstrate that this increased connectivity also expands the attack surface.
Critical infrastructures are increasingly defined by their exposure to modern cyber-physical threats, i.e., coordinated campaigns that blend intrusion, manipulation of control logic, and disruption of operational processes to trigger outages, unsafe states, or strategically significant impacts.
As these systems become more connected and data-driven, adversaries gain more paths to access and influence operational environments, raising both the likelihood and potential severity of deliberate, engineered disruption.

Industries have used \gls{cft} mechanisms to handle predictable crashes and hardware faults, relying on redundancy under benign failure assumptions~\cite{chintamaneni2002fault, ertugrul2002fault}.
However, the rise of intentional and adversarial behavior required stronger models, which renewed interest in \gls{bft}~\cite{nasreen2016study}.
Advances in distributed systems, networking, and computation capacity make these solutions practical, enabling systems to tolerate arbitrary and malicious faults.

Nonetheless, cyber-physical environments continue to exhibit unpredictable and adversarial behaviors that go beyond conventional \gls{bft} fault assumptions, potentially driving protocols into degraded or undefined states~\cite{huang2024security}.
Although organizations employ cyber ranges and attack labs to explore these conditions, their dependence on simulated datasets restricts fidelity, timing realism, and operational nuance, particularly in industrial systems~\cite{urias2018cyber}.

These limitations have contributed to the adoption of \glspl{dt}~\cite{grieves2016digital}: high-fidelity virtual replicas continuously fed with live telemetry.
By hosting or mirroring \gls{bft}-based components within \glspl{dt}—an approach already explored in prototypes for intrusion-tolerant SCADA and blockchain-backed \gls{dt} systems\cite{nogueira2018challenges, suhail2022blockchain}—organizations can safely experiment with alternative configurations, fault injections, and adversarial behaviors under conditions that reflect live operations.
In doing so, \glspl{dt} act as realistic sandboxes that help reveal latent weaknesses, cascading failures, and scenarios that could lead to halting, unrecoverable states, or compromise, thereby supporting more adaptive and operationally grounded resilience for \gls{bft}-enabled critical infrastructures.

Critically, existing fault injection approaches treat testing as a pre-deployment activity on static, synthetic testbeds, but industrial systems evolve, workloads shift, and adversaries adapt.
ByzTwin-Range reframes testing as a continuous operational loop: by feeding live telemetry into the twin, the system perpetually re-evaluates its own \gls{bft} guarantees under current real-world conditions.
This transforms the twin from a diagnostic tool into a self-hardening mechanism, one that can catch the timing windows and configuration drifts that only emerge under genuine operational load.

This paper introduces a novel framework that allows the prediction of potential faults, identifies the conditions in which they may arise, and provides mechanisms for handling these situations before they occur in reality.
This is achieved by employing a \gls{dt} and using real-world data, creating a cyber-range environment.


\subsection{Contributions}
This paper presents the following contributions:
\begin{itemize}
   \item \textbf{ByzTwin-Range, a dual-layer architecture integrating a \gls{bft} protocol with a high-fidelity \gls{dt}.} 
   A novel architecture where a production-grade \gls{bft}~\cite{HyperledgerFabric2024_SmartBFT, dong2025byzantine, hao2024doppel} deployment is conceptually mirrored within a \gls{dt}~\cite{jeremiah2024comprehensive}, enabling replication of protocol execution under real operational conditions.
   \item \textbf{A \gls{dt}–assisted engine for controlled Byzantine fault injection~\cite{soares2021zermia, pinto2023infrastructure}.}
   A fault-injection model is established that leverages the \gls{dt}'s live state to drive realistic Byzantine behaviors, including equivocation, inconsistent state reports, selective message loss, and targeted replica disruption, in ways that cannot be safely tested on production systems.
   \item \textbf{Automatic identification of \gls{bft} vulnerability regions and synchrony stress conditions.} 
   The design outlines how \gls{dt}-driven fault injection can expose worst-case timing windows, correlated failures, and physical–cyber interactions that challenge the liveness and safety guarantees of practical \glspl{bft}.
   \item \textbf{A methodology that bridges practical \gls{bft} deployments with next-generation cyber-range experimentation.}
   \gls{dt}-based environments can serve as \gls{bft}-aware cyber ranges, overcoming the limitations of static or purely simulated testbeds and enabling systematic evaluation of protocol behavior under realistic physical–cyber conditions.
\end{itemize}

The above contributions are architectural and methodological in nature; a proof-of-concept implementation to empirically validate the framework is planned as immediate future work.

\subsection{Organization}
The remainder of the paper is organized as follows. 
Section~\ref{sec:relatedwork} reviews the state of the art in \gls{bft} and \gls{dt}. 
Section~\ref{sec:arch} presents a detailed description of the ByzTwin-Range architecture. 
Section~\ref{sec:challenges} discusses the additional challenges associated with this research, as well as directions for future work.  
Finally, Section~\ref{sec:conclusion} concludes the paper and outlines directions for future work.

\section{State of the Art}\label{sec:relatedwork}


\glspl{dt} have been increasingly adopted for security evaluation in \gls{cps}, supporting safe replication of operational conditions for intrusion analysis, \gls{soc} workflows, and live-state safety monitoring~\cite{shitole2021real, dietz2020integrating, eckhart2018towards}.

Beyond security monitoring, \glspl{dt} have also been explored as cyber-range and fault-injection platforms to examine \gls{cps} behavior under controlled disturbances.
Flammini~\cite{flammini2021digital} proposed \glspl{dt} as run-time predictive models to support resilience assessments and ``What-if'' exploration in \gls{cps}.
Nguyen et al.~\cite{nguyen2022digital} introduced a Test and Simulation (TaS) tool that mirrors IoT deployments to provide higher-fidelity experimentation than traditional simulated testbeds. 
Larsen et al.~\cite{larsen2023fault} demonstrated the value of \gls{dt}-based co-simulation for fault injection in robotic \gls{cps}, enabling the exploration of timing deviations and component failures in a reproducible manner.

The intersection of \glspl{dt} with \gls{bft} has been explored less extensively; however, existing efforts indicate complementary directions.
Dettoni et al.~\cite{dettoni2013byzantine} proposed TwinBFT, where each physical replica is paired with a virtual counterpart to reduce the number of required physical nodes from $3f+1$ to $2f+1$, illustrating how virtual redundancy can support practical \gls{bft} deployments.
Sahal et al.~\cite{sahal2022blockchain} combined blockchain with \glspl{dt} to coordinate decentralized pandemic alerting, demonstrating the feasibility of using consensus mechanisms to synchronize \glspl{dt} across stakeholders.
Amiri et al.~\cite{amiri2024bedrock} introduced Bedrock, a unified platform for implementing and experimentally evaluating \gls{bft} protocols across different variants, with a focus on systematic analysis within distributed systems settings.
The tension between \gls{bft} and real-time constraints has also been explored in the real-time systems community. Loveless et al.~\cite{loveless2021igor} proposed eager execution to reduce the latency overhead of \gls{bft} State Machine Replication in safety-critical systems, while Böhm et al.~\cite{bohm2024tinybft} addressed \gls{bft} replication for resource-constrained embedded devices, both highlighting that timing predictability remains an open challenge when deploying \gls{bft} in cyber-physical environments.

These developments lay the foundations for a \gls{dt}-enhanced environment that supports practical and controlled \gls{bft} experimentation.

\section{System Architecture}\label{sec:arch}

This section addresses the assumptions and architecture design of ByzTwin-Range.

\subsection{Threat Model and Assumptions}
\begin{figure}[htbp]
\centerline{\includegraphics[width=\columnwidth]{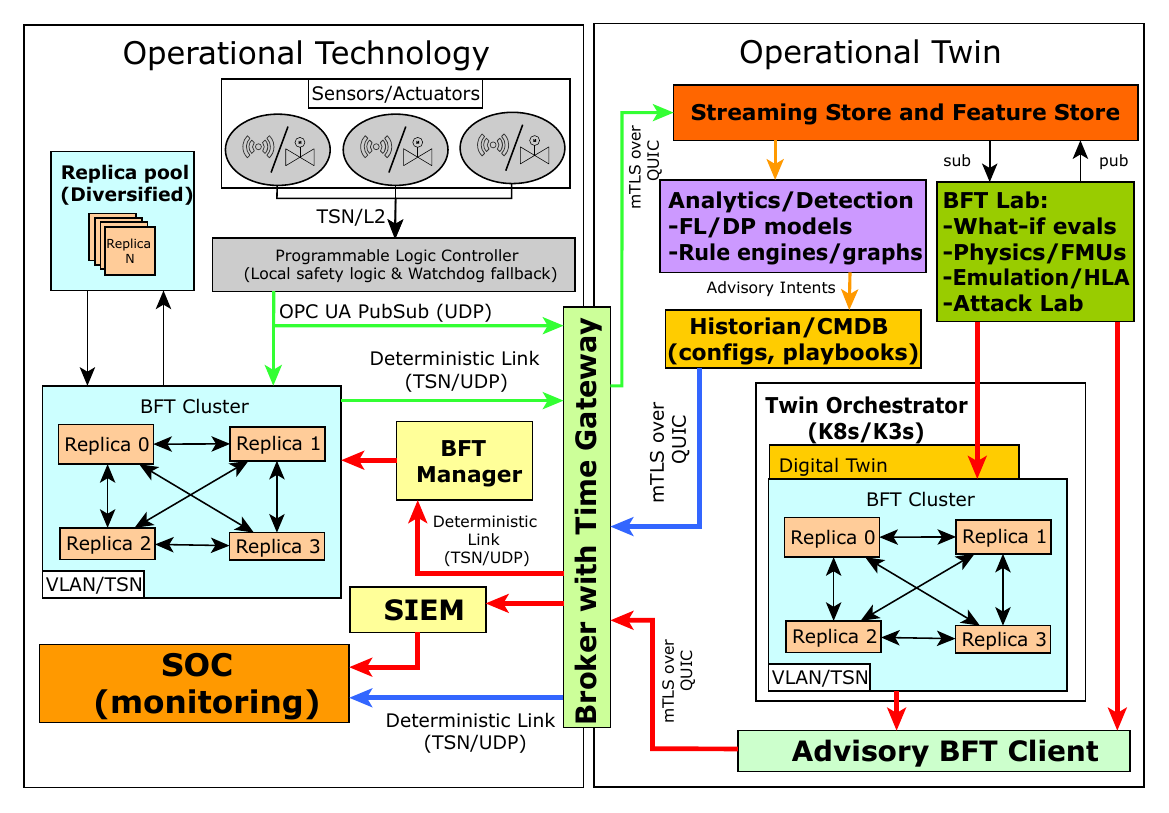}}
\caption{ByzTwin-Range's high-level topology for integrating \gls{bft} real-time control with a \gls{dt} simulation environment.}
\label{fig:arch}
\end{figure}

ByzTwin-Range assumes an adversary capable of compromising a bounded subset of replicas and influencing network timing, but not taking over the entire control environment. 
The system operates under partial synchrony: delays are finite but may be variable or adversarially perturbed.
Specifically, the \gls{ot} plane operates under strict real-time constraints enforced by \gls{tsn}, while the \gls{otw} and inter-layer communication channels operate under the partial synchrony model; Figure~\ref{fig:arch} reflects this boundary at the Broker and Time Gateway

At the protocol level, the \gls{bft} layer follows the standard model with up to $f$ Byzantine replicas among $3f{+}1$ replicas.
Byzantine nodes may equivocate, send inconsistent state, or manipulate message timing to induce false suspicions, deadline violations, or unnecessary view changes.

At the \gls{ot} layer, safety \glspl{plc} and watchdogs are assumed fail-safe, tolerating crashes or misbehaviour from the supervisory \gls{bft} layer and reverting to safe actuation on communication loss.
If this assumption is violated, for instance, in plants where safe states cannot be determined locally, the architecture would require a human-in-the-loop fallback, with the \gls{bft} layer demoted to advisory-only role until manual confirmation is obtained.
The physical process may exhibit noise and benign disturbances, but we do not consider a fully adversarial plant.

Within the \gls{otw}, the streaming store, time gateway, and orchestrator are treated as \gls{cft} components: they may fail or restart but do not present arbitrary behaviour (e.g., Byzantine).
Communication between the \gls{ot} and \gls{otw} uses mutual TLS mTLS) over QUIC with standard certificate-management practices.
Under this model, ByzTwin-Range concentrates on revealing protocol-driven and timing-driven vulnerabilities rather than defending against full-system compromise.

Figure \ref{fig:arch} illustrates the proposed ByzTwin-Range architecture, inspired by the TaS tool of Nguyen et al.~\cite{nguyen2022digital} and extended with additional cyber-range components for security analysis and controlled experimentation.
In this design, the \gls{dt} serves as the primary environment for evaluating and stress-testing the \gls{bft} \gls{otw} protocol, while the \gls{bft} \gls{ot} system itself remains deployed within the real operational environment.

The architecture is structured into two distinct planes: the \gls{ot} Plane and the \gls{otw} Plan.
The \gls{ot} Plane hosts the real industrial processes, controllers, and monitoring functions required to detect anomalies and assess potential threats.
The \gls{otw}, in contrast, provides an isolated execution environment where the \gls{dt} and its analytic components operate.
This separation enforces strict segmentation between the control layer and external attack surfaces, ensuring that experimentation, adversarial testing, and fault-injection activities in the \gls{otw} cannot compromise the operational system.

The \gls{dt} continuously mirrors live operational data, allowing safe validation of \gls{bft} behaviour, misconfigurations, and fault or attack scenarios without disrupting the physical system.
The insights produced by the twin are then fed back into the \gls{ot} to support adaptation, tuning, and resilience analysis.
This integration is enabled through established technologies, such as \gls{opc}, Kubernetes, and the \gls{hla}, which collectively ensure interoperability, scalability, and high-fidelity synchronisation between the physical system and its virtual counterpart.

\subsection{Operational Technology (OT)}

In this architecture, we consider a real-world \gls{cps} where sensors and actuators continuously send data to a \gls{plc}. 
The \gls{plc} executes local safety logic, maintains watchdog fallbacks, and acts as a real-time industrial controller at the edge of the \gls{ot} environment. 


To ensure resilience, the \gls{plc} includes embedded safety logic and watchdog mechanisms that enforce fail-safe actions if communication fails, the supervisory \gls{bft} layer malfunctions, or expected cycle times are exceeded.






The \gls{plc} performs low-level actuation, while \gls{bft} nodes coordinate higher-level supervisory decisions, backed by a diversified replica pool that allows compromised, unhealthy, or maintenance-bound replicas to be removed and replaced without disrupting system operation.

All components communicate over \gls{tsn2}, providing deterministic, low-latency Layer-2 Ethernet connectivity with bounded jitter, traffic prioritization, micro-segmentation, and strict time synchronization. 

Communication with the \gls{bft} \gls{ot} layer uses \gls{opc} Publish–Subscribe over UDP, enabling deterministic, low-latency, multicast-friendly telemetry for both the \gls{bft} controllers and the \gls{otw} environment; internally, \gls{bft} replicas interconnect via \gls{vlan} to ensure real-time determinism, traffic isolation, and operational safety. 

Data then flows through a deterministic \gls{tsn}-over-UDP link to the Broker and Time Gateway, which provide temporal and semantic coordination between the \gls{ot} plane and the \gls{otw}.

\subsection{Broker and Time Gateway}
The objective of this subsystem is to ensure temporal alignment, secure mediation, and event traceability across the \gls{ot}, \gls{otw}, and \gls{bft} layers. 
It achieves this through three main functions: (i) temporal alignment and deterministic ordering via the Time Gateway, (ii) secure and decoupled topic-based mediation through the Broker, and (iii) event provenance, logging, and reproducible replay across the system.

The first objective is achieved by the Time Gateway.
Since \gls{ot} components operate on fixed real-time cycles, \gls{otw} simulations may run at varying speeds, and \gls{bft} protocols depend on timestamps and quorum deadlines, synchronization across these time domains is essential.
The Time Gateway aligns \gls{ot} timestamps with \gls{otw} simulation time, converts between real, logical, and simulated time, enforces correct event ordering with delay annotation, and provides stable clocking for \gls{bft} state logging and replay.
This ensures that the \gls{otw} receives causally consistent data, enabling accurate reconstruction of real-world scenarios.

The second objective is accomplished through the Broker.
As \gls{ot} networks cannot expose raw \gls{plc} or safety-critical streams directly, the Broker serves as a controlled publish/subscribe layer between \gls{plc} publishers and \gls{otw}/\gls{bft} consumers.
It normalizes protocols (e.g., converting \gls{opc} PubSub to MQTT, Kafka, \gls{hla}, or \gls{fmu} formats), enforces security policies such as mTLS over QUIC, role-based access control, and message signing, and decouples \gls{ot} timing and load from the \gls{otw} environment.
This allows \gls{otw} workers, simulation nodes, and \gls{bft}-analysis modules to subscribe only to the necessary streams without impacting real-time \gls{ot} behavior.


The third objective is fulfilled jointly by the Broker and Time Gateway: they assign canonical timestamps, maintain cross-layer traceability, and enable reproducible replay of operational sequences, allowing \gls{otw}-detected anomalies to be mapped back to real events.
Telemetry is then forwarded to a \gls{siem} and \gls{soc} for continuous monitoring.

\subsection{Operational Twin (OTw)}
The objective of the \gls{otw} is to create a \gls{dt} of the \gls{bft} system deployed in the \gls{cps}.
It simulates and evaluates potential configurations that may lead to faults, whether from normal behavior or adversarial actions, allowing analysts to take preventive measures.
This is achieved through ``What-if'' tests using real-world data, \glspl{fmu}, \gls{hla}-based emulations, and attack scenarios. 

To achieve these objectives, the \gls{otw} forwards real-world telemetry and event data from the \gls{ot} using mTLS over QUIC, ensuring secure, authenticated, low-latency communication.
The data is ingested by a streaming store, which acts as an append-only, time-ordered log, enabling buffering, replay, and fan-out to multiple consumers without disrupting the live system.

A feature store maintains curated, versioned Machine Learning features to support \gls{fl} and \gls{dp} models, ensuring consistency in model training and serving while preventing train–serve skew.

Concretely, \gls{fl} models trained across replica nodes detect anomalous timing patterns and behavioural deviations, such as abnormal view-change frequencies or quorum deadline violations, without centralising raw operational data, while \gls{dp} guarantees that model updates cannot leak sensitive process state.

The \gls{bft} Lab uses the data and feature streams to run ``What-if'' evaluations with \glspl{fmu} and \gls{hla}-based emulations, simulating different system configurations and their effects under normal and adversarial conditions. 
\glspl{fmu} encapsulate component-level models for co-simulation, and \gls{hla} federates these models into a distributed ``twin federation,'' enabling synchronization across simulators.
This allows ByzTwin-Range to recreate realistic multi-domain scenarios while isolating experiments from production systems.

Simulation workloads are managed by Kubernetes (K8s) or K3s, enabling repeatability, scaling, and controlled testing of new \gls{bft} configurations in isolation from the live system.

Validated outcomes, such as advisories, anomaly detections, and failure patterns, are sent to the Broker via mTLS over QUIC.
The advisory channel is intentionally asymmetric: the \gls{otw} issues read-only intents to the \gls{bft} Manager, which retains sole authority over actuation decisions.
This design ensures that a compromised or manipulated twin cannot directly alter operational behaviour, preserving the security boundary between experimentation and production.
The Broker distributes these to the \gls{bft} \gls{ot} cluster, forwards events to the \gls{siem}/\gls{soc}, and republishes simulation outputs to the streaming and feature stores for further analysis and model support.

Lastly, verified configurations, playbooks, and supporting evidence are stored in the Historian/CMDB, creating a knowledge base that strengthens future deployments, accelerates onboarding, and enhances system resilience.

\section{Additional Challenges/Future Work}\label{sec:challenges}

Despite its potential, ByzTwin-Range raises several challenges. 
First, aligning the worst-case latency of \gls{bft} with the deterministic timing of industrial control remains difficult, especially under view changes or adversarial network conditions.
Future work will integrate runtime timing monitors and twin-assisted prediction to ensure consensus decisions stay within safety-critical deadlines.

Second, maintaining a consistent state between the operational system and its \gls{dt} counterpart is non-trivial. 
Small divergences in timing, noise, or nondeterminism may accumulate and bias experiment results. 
Lightweight snapshot/replay mechanisms, along with deterministic event ordering, are necessary to prevent long-term drift.

Third, scaling high-fidelity co-simulations (\gls{fmu}/\gls{hla}) to realistic industrial deployments introduces computational overhead and orchestration complexity. 
Exploring adaptive-fidelity models and more efficient scheduling strategies will be important to support continuous experimentation.

Lastly, extending the fault-injection engine to cover coordinated, cross-layer, and protocol-aware adversarial behaviors, and validating their physical impact through the \gls{dt}, constitutes key future work.


\section{Conclusion}\label{sec:conclusion}
In conclusion, ByzTwin-Range introduces a continuous, live-data-driven approach for testing and hardening \gls{bft} deployments in cyber-physical environments. 
The architecture is grounded in proven industry-standard technologies, supporting feasibility and compatibility with existing operational workflows. 
The next step is a proof-of-concept to validate the framework under real-time performance conditions.

\bibliographystyle{elsarticle-num}
\bibliography{bibliography.bib}

\end{document}